\documentclass[prl,showpacs,superscriptaddress,twocolumn]{revtex4}
\usepackage{texdraw}
\usepackage{amsmath}
\usepackage{amsfonts}
\usepackage{amssymb}
\usepackage{graphicx}
\usepackage{array,tabularx}

\def\tu{t_{u}}

\begin{document}
\title{Unjamming dynamics: the micromechanics of a seismic fault model}
\author{Massimo Pica Ciamarra}\email[]{picaciamarra@na.infn.it}
\affiliation{CNR--SPIN, Department of Physical Sciences, University of Naples 
Federico II, 80126 Napoli, Italy.}
\homepage{http://smcs.na.infn.it}
\author{Eugenio Lippiello}
\affiliation{Dep. of Environmental Sciences and CNISM,
Second University of Naples, 81100 Caserta, Italy}
\author{Cataldo Godano}
\affiliation{Dep. of Environmental Sciences and CNISM,
Second University of Naples, 81100 Caserta, Italy}
\author{Lucilla de Arcangelis}
\affiliation{Dep. of Information Engineering and CNISM,
Second University of Naples, 81031 Aversa (CE), Italy}
\affiliation{Institute for Building Materials, Schafmattstr. 6, ETH, 8093 Z\"urich, CH}

\begin{abstract}
The unjamming transition of granular systems is investigated in a seismic fault model 
via three dimensional Molecular Dynamics simulations. 
A  two--time force--force correlation function, and a susceptibility related to the system response to pressure 
changes, allow to characterize the stick--slip dynamics, consisting in large slips 
and microslips leading to creep motion.
The correlation function unveils the micromechanical changes 
occurring both during microslips and slips.
The susceptibility encodes the magnitude of the incoming microslip.
\end{abstract}

\maketitle

\begin{figure}[t!]
\includegraphics{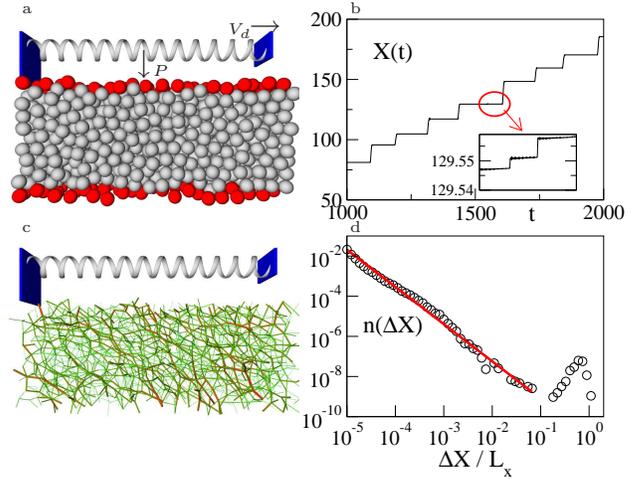}
\caption{\label{fig:model}
(Color online) (a) The system consists of granular particles (light grains)
confined between two rigid plates (dark grains) at constant pressure. 
The top plate is driven via a spring mechanism, where an extremum of the spring is attached
to the plate and the other is pulled at constant velocity.
See the supplementary materials available online for an animation of a slip event in the real and in the force space~\cite{epaps}.
(b) Time evolution of the top plate position $X(t)$. 
The inset indicates that between two large slips there are many microslips. 
(c) Configuration of normal force network. Stronger forces correspond to thicker and darker lines.
(d) Slip size distribution for slips and microslips.
}
\end{figure}

In a number of industrial processes and natural phenomena, such as earthquakes 
or landslides, disordered solid granular systems start to flow. This solid-to-liquid transition, 
known as unjamming, occurs either on decreasing the confining pressure $P$,
or increasing the applied shear stress $\sigma$. Understanding the properties of this
transition is a big challenge due to the absence of an established theoretical 
framework for granular materials. 
A proposed analogy with the glass transition~\cite{Liu98} of thermal systems 
has recently triggered the study of the jamming transition via numerical investigations
of systems of soft frictionless particles at zero applied shear stress~\cite{Ohern03}, 
where the only control parameter is the pressure (or the density). 
As the unjamming transition is approached by decreasing the confining pressure, the vibrational spectrum
develops an excess of low frequency modes, known as soft--modes, leading
to the identification of a length scale which diverges on unjamming~\cite{Silbert05}. 
This length scale is related to the emergence of an increasingly heterogeneous response as the system moves towards the 
transition~\cite{Silbert05}. 
A different approach to the study of the unjamming transition has been followed in 
a two dimensional numerical study~\cite{Aharonov2004} and in a number 
of experiments~\cite{Nasuno1997,Gollub2003,Dalton,Nori2006}, where the applied shear stress is controlled via a spring mechanism, as the one in Fig.~\ref{fig:model}a. 
A stick--slip motion characterized by a complex slip size statistics~\cite{Dalton} is recovered at high confining pressures $P$ and small driving velocities $V_d$.
This stick-slip dynamics is altered by the presence of noise~\cite{Zapperi2009}.
Analogous results have been found at fixed strain rate~\cite{Johnson2005,hans}.

In this Letter we tackle this problem via three dimensional Molecular Dynamics simulations
of a model of a seismic fault (Fig.~\ref{fig:model}a), where grains play the role of the gouge~\cite{Nasuno1997,Gollub2003,Dalton,Nori2006,Johnson2005,hans}.
Numerical details are given in~\cite{nota,erice}.
The micromecanical mechanisms leading to the transition are analyzed at a level of spatial and temporal resolution
not considered before.


{\it Stick--slip dynamics -- } 
For the investigated values of the parameters~\cite{nota}, the system is characterized by stick--slip dynamics,
which we analyse considering that a slip begins and ends when
the velocity of the top plate becomes, respectively, larger and smaller than a small enough threshold.
We measure the displacements $\Delta X$ of the top plate due to slips, and compute their distribution $n(\Delta X)$ (Fig.~\ref{fig:model}d).
For slips smaller than $\simeq 0.1 L_x$, where $L_x$ is the system length, the distribution follows a power law, $n(\Delta X) \propto \Delta X^{-\beta}$ with $\beta \simeq 1.85$, in agreement with experimental values for earthquakes~\cite{kan}. 
Larger slips are almost periodic in time and roughly follow a lognormal distribution with a characteristic size $\Delta X \simeq 0.6 L_x$.
Summarizing, the dynamics consists in the occurrence of almost periodic large events, here called {\it slips},
and of creep motion characterized by smaller events, here called {\it  microslips},
in agreement with previous experiments~\cite{Dalton,Johnson2005}.
\begin{figure}[t!]
\begin{center}
\includegraphics*[scale=0.3]{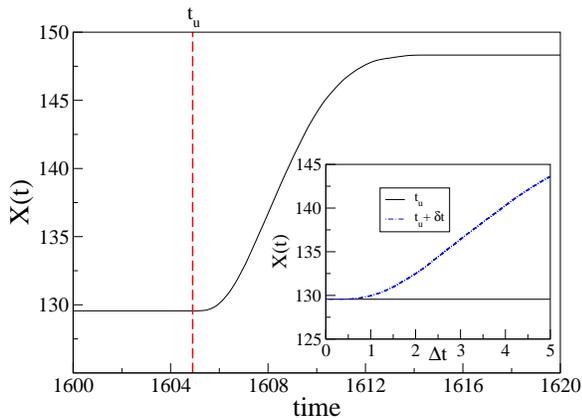}
\end{center}
\caption{\label{fig:Xt}
(Color online) The top plate position in a slip event $X(t)$. 
The vertical dashed line indicates the unjamming time $\tu$. 
Inset: Top plate position for two system replicas, which evolve with $V_d = 0$. 
The replica made at time $\tu$ remains in the jammed configuration (continuous line), whereas the one made at time 
$\tu + \delta t$ slips (dashed line). 
}
\end{figure}

{\it Onset of a slip -- }
To understand the mechanisms acting at the onset of a slip, we need to the identify its precise starting time $\tu$.
Here we describe the analysis performed on the particular slip occurring at time $t \simeq 1600$. 
We consider a replica of the system at time $t$, and follow its time evolution at zero driving velocity $V_d = 0$.
If the replica made at time $t$ resists to the applied stress, then $t \le \tu$. Conversely, if a slip is observed, $t > \tu$.
We define $\tu$ as the largest time where no slip occurs, and we identify it (one for each slip)
with an accuracy equal to the time step of integration of  the equation of motion $\delta t$~\cite{nota} (Fig.~\ref{fig:Xt}). 
This procedure is equivalent to a quasi-static simulation~\cite{Barrat2009} around the un-jamming time $\tu$ and gives 
the value of the shear stress above which the system starts to flow.

We have performed a number of checks which suggest that no structural changes occur at $t_u$.
For instance, the comparison of the state of the system at time $\tu$ with the one at shortly earlier and later 
times, shows that no contact breaks at $\tu$.  We have also considered the distribution of the parameter
$\lambda = |{\bf f}_t|/\mu|{\bf f}_n|$ where ${\bf f}_t$ and ${\bf f}_n$ are the tangential and the normal forces.
When $\lambda > 1$ a contact breaks as the Coulomb condition is violated. The maximum of the probability distribution $P(\lambda)$ gradually moves toward $1$ as $\tu$ is approached, indicating the weakening of the solid~\cite{Aharonov2004}. 
However, neither at  $\tu$ the number of contacts with $\lambda \simeq 1$ overcomes a given fraction, nor they appear to be spatially organized.
The absence of structural changes at $\tu$ supports a scenario in which the system is located in an energy minimum which slowly becomes an inflection point at time $\tu$, and therefore the smallest eigenvalue of the dynamical matrix continuously decreases to zero ~\cite{Maloney2006,Mcnamara2009}.

\begin{figure}[t!]
\begin{center}
\includegraphics*[scale=0.3]{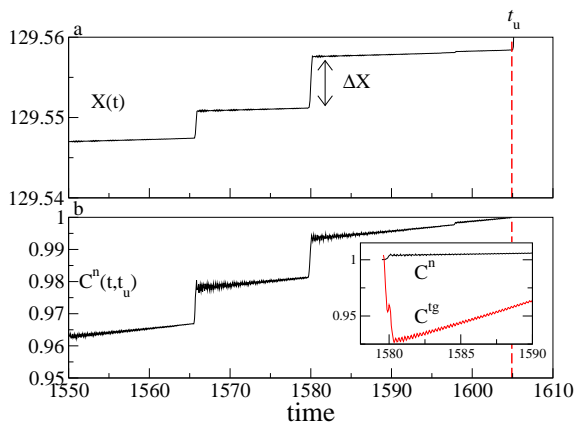}
\end{center}
\caption{\label{fig:microslip}
(Color online) Time evolution of the top plate position (a) and of the correlation 
function of the normal forces $C^n(t,\tu)$ (b). Dashed lines identify the unjamming time $\tu$.
The inset show $C^n$ and $C^{tg}$ across a microslip at $t = 1579.5$.
}
\end{figure}

{\it Evolution in the force space --} 
When the system sticks and the shear stress increases, no macroscopic motion is observed. 
However, the system microscopically changes since it sustains an increasing shear stress. 
The time evolution of the top plate position $X(t)$ (Fig.~\ref{fig:microslip}a), consists 
in an elastic deformation, where $X(t)$ increases very slowly in time, interrupted by sudden 
microslips. To characterize the evolution of the system in the force space, we introduce a two time 
force-force correlation function for the normal forces, defined as
\begin{equation}
\label{eq:corrf}
C^{n}(t_0,t) = \frac{\sum_{ij} |{\bf f}_{ij}^{n}(t_0)||{\bf f}_{ij}^{n}(t)|}{\sum_{ij} |{\bf f}_{ij}^{n}(t_0)|^2},
\end{equation}
where the sum running over all couples of particles $(i,j)$ corresponds to a spatial average.
An equivalent definition holds for the correlation of tangential forces, $C^{tg}$.
Being interested in the unjamming transition of the slip event shown in Fig.~\ref{fig:Xt},
here we fix $t_0 = \tu$, and consider the evolution of the correlation function $C^n(t_u,t)$ for earlier times, $t < \tu$ (Fig.~\ref{fig:microslip}b). 
The force correlation function $C^n$ (and  $C^{tg}$, not shown) exhibits small jumps in correspondence of microslips (Fig.~\ref{fig:microslip}a), revealing the unusual occurrence of bursts in the reorganization of the force network.
During these bursts, the energy due to the tangential interaction decreases, 
whereas the one due to the normal interaction increases. 
A possible interpretation is in terms of a two force network scenario, in which
the applied stress $\sigma$ is supported by a stress $\sigma_n$ due to the normal force network, and by 
a stress $\sigma_t$ due to the tangential forces, $\sigma = \sigma_n + \sigma_t$. 
In a burst, few contacts break, leading to a decrease of $\sigma_t$, $\sigma_t \to \sigma_t - \delta\sigma$. 
A microscopic slips is observed since the normal forces quickly adapt and succeed 
in sustaining the applied stress, $\sigma_n \to \sigma_n + \delta\sigma$.
This scenario is supported by the inset of Fig.~\ref{fig:microslip}b, which shows both $C^{n}$ and $C^{tg}$ across 
a microslip, with $t_0$ its starting time. $C^{n}$ slightly increases and overcomes $1$, while $C^{tg}$ exhibits a sharp drop due to the breaking of several contacts.

\begin{figure}[t!]
\begin{center}
\includegraphics*[scale=0.3]{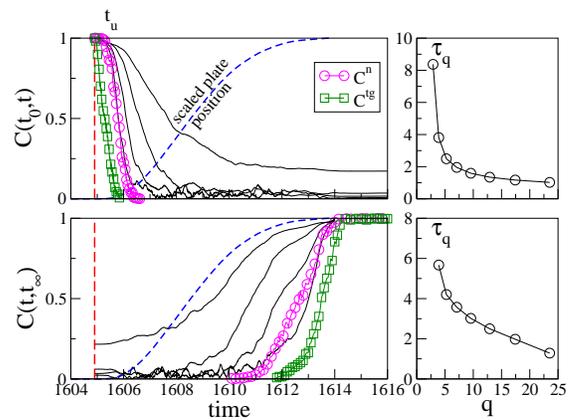}
\end{center}
\caption{\label{fig:corr_forces}
(Color online) Correlation functions at the unjamming (upper panels) and at the subsequent jamming transition (lower panels).
The left panels show the normal and tangential force correlation functions $C^{n,tg}(t_0,t)$ (symbols)
during a slip event. 
The dashed line shows the time evolution of the plate position scaled between $0$ and $1$.
The plain lines show the self-scattering function $F(q,t)$ for $q = 3,5,9,17$.
The corresponding $\tau_q$ are shown in the right panels.
$t_0$ is fixed equal to $\tu$ for the unjamming transition, and to $t_0 = 1614$ for the subsequent jamming.
}
\end{figure}

The force correlation function (Eq.~\ref{eq:corrf}) gives also insights into the system evolution during a slip.
In Fig.~\ref{fig:corr_forces} we plot $C^{n,tg}(t_u,t)$ for $t > \tu$, and for comparison the scaled top plate position 
$(X(t)-X(\tu))/(X(t_\infty) - X(\tu))$, where $t_\infty$ is a time 
following the slip event, whose precise value does not influence our results.
We first notice that the forces evolve on a timescale much shorter than the plate motion. 
For instance the correlation functions reach the value $0.1$, 
denoting an almost complete relaxation, when the top plate moved only by 
$10\%$ of its total displacement. 
The presence of different time scales in the relaxation process is evidenced by the self-scattering 
correlation function $F(q,t) = \frac{1}{N} \left| \sum_j \exp[i{\bf q}\cdot({\bf r}_j(t)-{\bf r}_j(t_u))] \right|^2$,
where ${\bf r}_j$ represents the position of the $j$th particle.
Since the system is sheared along $x$ and confined along $z$, we have considered wave vectors along $y$, ${\bf q} = (0,q,0)$. For large $q$, $F(q,t)$ probes small scale relaxation, and coincides with $C^{n}(t_u,t)$, as shown in Fig.~\ref{fig:corr_forces}. Conversely, at small $q$, $F(q,t)$ relaxes on a time scale comparable to that of the upper plate motion. 
The relaxation time $\tau_q$,  $F(q,\tau_q) = 1/e$, indeed increases as $q$ decreases. 
Moreover, tangential forces decorrelate before normal ones. 
This can be explained considering the unjamming transition as a buckling-like 
instability of the chains of large normal forces, which are sustained by weaker tangential contacts. 
When the weaker sustaining contacts break, either the normal forces adapt to sustain the extra
load, leading to a microslips, or a buckling-like instability occurs, giving rise to a slip.
The same quantities can be used to 
investigate the subsequent jamming transition. To this end, the force network 
at time $t$ is compared with the force network after the slip event studying $C^{n,tg}(t,t_\infty)$.
Tangential forces correlate after normal ones, and makes the force network stable.

\begin{figure}[t!]
\begin{center}
\includegraphics*[scale=0.3]{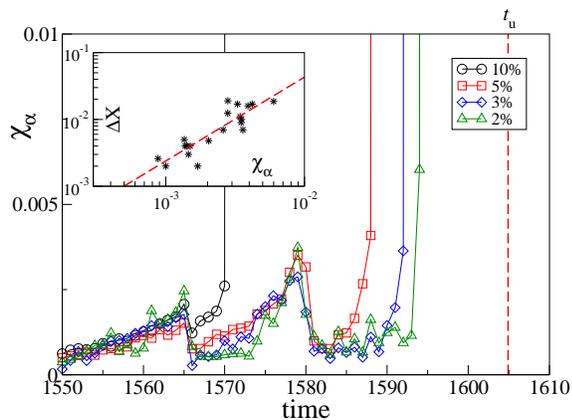}
\end{center}
\caption{\label{fig:length}
(Color online) Time evolution of the susceptibility $\chi_\alpha$ for different $\alpha$.
(Inset) Size of a microslip $\Delta X$ versus the corresponding value  $\chi_\alpha$. 
The straight line is $\chi_\alpha^b$, with $b \simeq 1.2$.
}
\end{figure}

{\it Response to perturbations: slips versus microslips -- }
The correlation length $\xi$ in equilibrium systems is measured from the response to an external perturbation. 
The susceptibility, for instance, scales like $\xi^{2-\eta}$ near critical points, where $\eta$
is the correlation function critical exponent. 
Here we measure the response of the system to a pressure change at the onset of
both slips and microslips.
More precisely, at each time $t$ we stop the external drive, setting $V_d = 0$, and 
introduce a perturbation in the external pressure $P$, 
fixed to $P' = P(1-\alpha)$ for a time interval $\delta t_{perb} = 0.1$. 
The response $\chi_\alpha$ at time $t$ is defined as
\begin{equation}
\label{eq:xi}
\chi_\alpha(t) = 
\frac{1}{\alpha P}
\lim_{\tau \to \infty}   \left[ \frac{1}{N} 
\sum_i ({\bf r}_i^\alpha(t+\tau)-{\bf r}_i^0(t+\tau))^2 \right]^{1/2}
\end{equation}
where $\{{\bf r}^\alpha_i\}$ and $\{{\bf r}^0_i\}$ are the asymptotic states of 
the perturbed of the unperturbed systems. In the unjammed phase, the susceptibility is divergent.
In the jammed phase, it measures the size of the region of correlated particles that respond to the
external perturbation, providing an estimate of the correlation length.
It is a static quantity since the time $t$ only indicates the instant at which the perturbation is applied.
$\chi_\alpha$ can be also defined without setting $V_d = 0$, provided that the characteristic response time of the system is much smaller than the timescale over which the applied stress varies.
For a wide range of $\alpha$, the response of the system at times far from $\tu$ 
is linear in the perturbation, since $\chi_\alpha$ does not depend on $\alpha$ (Fig.~\ref{fig:length}).  
In particular, $\chi_\alpha$ gradually increases in time and drops in correspondence to microslips. 
The inset shows that the microslip size $\Delta X$
depends on $\chi_\alpha$ evaluated just before the slip, as $\Delta X  \simeq \chi_\alpha^b$.
This indicates that the size of a microslip is already encoded in the system state.
In fact, considering that $\chi_\alpha$ increases on approaching a microslip, the measured $\chi_\alpha$
provides a lower bound for the magnitude of the incoming event.

As the unjamming time is approached, the response is no longer linear. $\chi_\alpha$ remains roughly constant until it abruptly increases at a time which depends on $\alpha$, the sooner the greater $\alpha$. This increase is consistent with a power-law divergence. Accordingly at each time, namely at each value of the applied shear stress, there is a minimum value 
of the perturbation intensity for slip triggering. This behavior is in line with the existence of a minimum threshold amplitude in the deformation associated with seismic waves for earthquake remote triggering~\cite{gom}.
A difference between slips and microslips is in how different particles contribute to $\chi_\alpha$ in the sum of Eq.~\ref{eq:xi}. Indeed, these contributions are very similar for microslips, and highly heterogeneous for slips. The presence of an heterogeneous response is consistent with previous numerical results found at $\sigma = 0$~\cite{Silbert05}.

In conclusion, the absence of precise structural changes at the unjamming time,
and the bursts observed in the prior dynamics,
suggest that the increasing external stress progressively 
modifies the underlying energy landscape. 

At each time the system is in an equilibrium position which can be
seen as local energy minimum of an effective energy landscape which depends on the applied shear stress.
Microslips occur when the local energy minimum flattens down as the applied shear stress increases,
letting the system fall in a neighbor minimum. If there are no close 
minima, a slip occurs, and the system jumps to a far away configuration.
The deforming energy landscape picture also suggests that soft-modes
could be found not only when unjamming is approached at $\sigma = 0$ by decreasing 
the volume fraction $\phi$~\cite{Silbert05}, but also~\cite{PicaCiamarra09} when 
unjamming is approached increasing $\sigma$.

\begin{acknowledgments}
We thank A. Coniglio for helpful discussions and the University of Naples Scope grid project, CINECA and CASPUR for computer resources.
\end{acknowledgments}


\begin{thebibliography}{40}
\bibitem{nota}{We consider $N = 10^3$ particles of mass $m$ and diameter $D$ confined in the $z$-direction by two 
rigid rough plates with size $L_x = 20D$, $L_y = 5D$. Periodic boundary conditions are imposed along $x$ and $y$.
Particles interact in the normal direction via the standard linear--spring 
dashpot model~\cite{erice}, characterized by an elastic constant $k_n$ and by a restitution coefficient 
$e = 0.88$. In the tangential direction, we use the spring model developed in Ref.~\cite{Cundall} with
elastic constant $k_t = 2/7k_n$ and coefficient of friction $\mu = 0.1$. 
Lengths, masses, times and stress are expressed in units of $D$, $m$, $\sqrt{k_n/m}$ 
and $k_n/D$, respectively. The confining pressure is $P = 100$ and $k_{\rm d} V_{\rm d} = 10$, 
where $k_{\rm d}$ is the elastic constant of the driving spring. The time step of integration of the equation of motion is $\delta t = 2~10^{-5}$. The slipping velocity threshold is $v_t = 10^{-5}$.}
\end{thebibliography}
\end{document}